\documentclass[pra,twocolumn,preprintnumbers,amsmath,amssymb,superscriptaddress]{revtex4}

\usepackage{color}
\usepackage{graphicx}% Include figure files
\usepackage{dcolumn}% Align table columns on decimal point
\usepackage{bm}% bold math
\usepackage{hyperref}
\usepackage{mathrsfs}
\usepackage{soul}
\usepackage{ulem}
\pdfoutput=1

\begin{document}
	
	%\preprint{APS/123-QED}

	\title{Entanglement generation and detection in split exciton-polariton condensates}	
	
	\author{Jingyan Feng}
	\affiliation{State Key Laboratory of Precision Spectroscopy, School of Physical and Material Sciences, East China Normal University, Shanghai 200062, China}
	
	\author{Hui Li}
	\affiliation{State Key Laboratory of Precision Spectroscopy, School of Physical and Material Sciences, East China Normal University, Shanghai 200062, China}
	
	\author{Zheng Sun}
	\affiliation{State Key Laboratory of Precision Spectroscopy, School of Physical and Material Sciences, East China Normal University, Shanghai 200062, China}
	
	\author{Tim Byrnes}
	\email{tim.byrnes@nyu.edu}
	\affiliation{New York University Shanghai, 567 West Yangsi Road, Shanghai, 200126, China; NYU-ECNU Institute of Physics at NYU Shanghai, 3663 Zhongshan Road North, Shanghai 200062, China; Shanghai Frontiers Science Center of Artificial Intelligence and Deep Learning, NYU Shanghai, 567 West Yangsi Road, Shanghai, 200126, China.}
	\affiliation{State Key Laboratory of Precision Spectroscopy, School of Physical and Material Sciences, East China Normal University, Shanghai 200062, China}
	\affiliation{Center for Quantum and Topological Systems (CQTS), NYUAD Research Institute, New York University Abu Dhabi, UAE.}
	\affiliation{Department of Physics, New York University, New York, NY 10003, USA}

	\date{\today}% It is always \today, today,
	
	\begin{abstract}
		We propose a method of generating and detecting entanglement in two spatially separated exciton-polariton Bose-Einstein condensates (BECs) at steady-state. In our scheme we first create a spinor polariton BEC, such that steady-state squeezing is obtained under a one-axis twisting interaction.  Then the condensate is split either physically or virtually, which results in entanglement generated between the two parts. A virtual split means that the condensate is not physically split, but its near-field image is divided into two parts and the spin correlations are deduced from polarization measurements in each half. We theoretically model and examine logarithmic negativity criterion and several correlation-based criteria to show that entanglement exists under experimentally achievable parameters.
	\end{abstract}
	
	%\pacs{03.75.Dg, 37.25.+k, 03.75.Mn}% PACS, the Physics and Astronomy
	
	\maketitle
	
	\section{Introduction}\label{i}
	Entanglement is a central property of quantum physics that distinguishes it from classical physics
	\cite{Vedral97,Vedral14}, and is considered an essential resource for applications such as quantum information \cite{Bennett93,Lee02}, quantum cryptography \cite{Gisin02,Yin2020,Ekert91} and quantum metrology \cite{Giovannetti2011,pezze18}. Entangled states have already been achieved at the macroscopic scale, and in systems such as atomic ensembles \cite{Krauter13} and mechanical resonators \cite{Kotler21}. Several experiments realized the generation of entanglement and other quantum correlations between the atoms of a single BEC cloud \cite{Lange2018,Kunkel2018,Fadel18,Schmied16}, which have been proposed for several applications \cite{Berrada13,Esteve08}. A well-known platform of creating BECs is with suitably structured semiconductor systems supporting exciton-polaritons. Exciton-polaritons are a superposition of an exciton (an electron-hole bound pair) and a cavity photon, and form a bosonic quasiparticle \cite{Deng10,Kasprzak06,Jonathan11,Byrnes2014}. The coupling between the exciton and photon results in an extremely light mass for the exciton-polaritons \cite{Deng02,Byrnes2014}, allowing for the possibility of realizing BECs \cite{Kasprzak06,Deng02,Balili07}. One of the advantages of polariton BECs is that they can be experimentally implemented at higher temperatures, even at room temperatures, by using materials such as GaN, ZnO. \cite{Christopoulos07,Baumberg08,kena10,Guillet11,Plumhof14,Fei22}. This makes the polariton system attractive for future technological applications, as they would not require bulky cryogenic apparatus.

	Currently, entanglement in spatially separate Bose-Einstein condensates, of any species (atomic, polaritonic, or otherwise) is yet to be observed. However, experimental demonstration of generating and detecting entanglement between spatially separated regions of a single BEC has been achieved \cite{Lange2018,Kunkel2018,Fadel18}. In these works, entanglement was first created between the atoms on a single $^{87}$Rb atomic BEC, using methods such as state-dependent forces, spin-nematic squeezing and spin-changing collisions. Then by using a magnified near-field image of the single atomic BEC, two different spatial regions of the same BEC were examined for correlations. It was shown that the entanglement can be detected after releasing the atomic gases from the traps. While the splitting process is only virtual, and not physically separated, this constitutes the first step to showing that BECs can be put in an entangled state. Numerous theoretical proposals have been made for generating entanglement in two completely separate atomic BECs \cite{Treutlein06,yumang19,kitzinger20,Idlas16,Pyrkov13,Pettersson17,Abdelrahman14,Rosseau14,Hussain14}. For polariton condensates, to date, no reports of detection of entanglement in a single or multiple polariton condensates have been made. The virtual splitting procedure may be an excellent first candidate for observing such entanglement. As shown in Ref. \cite{yumang19}, a physical or virtual split gives identical results in terms of entanglement, and extensions of the approach to completely separate BECs may be performed in the future. A very promising first candidate for entanglement between spatially separated BECs would be polariton BECs, which can be easily manipulated \cite{Kim08,Estrecho21,yingjie22}.

\begin{figure}
		\includegraphics[width=\columnwidth]{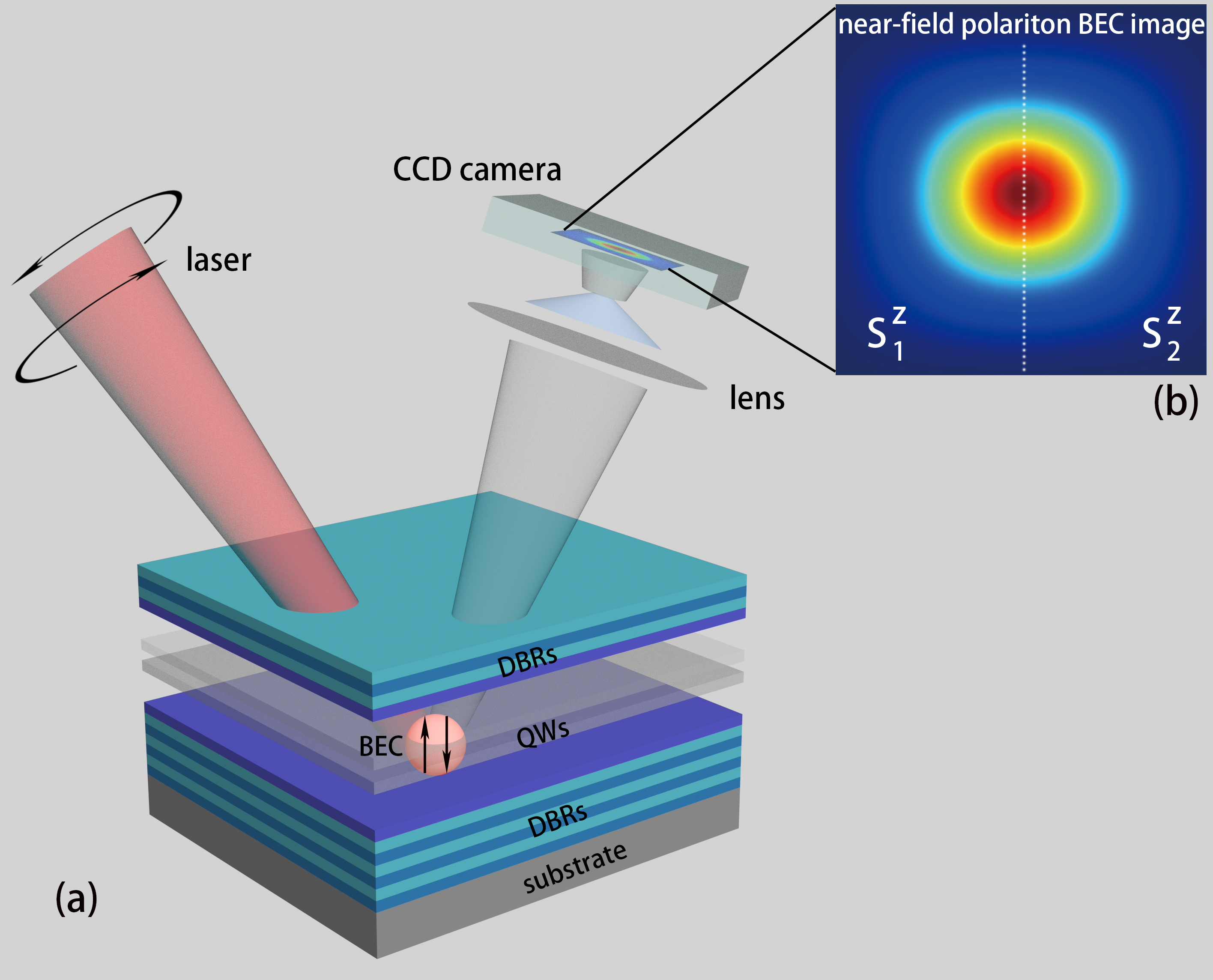}
		\caption{The experimental setup for our system. (a) A spinor exciton-polariton BEC forms in the QWs generated by pump laser. The spinor BEC is formed from the spin components of the polaritons, and are excited by applying a laser of suitable polarization (both clockwise and anti-clockwise circular polarization) to excite equal populations of the spins. The photon component of the polaritons leaks through semiconductor quantum microcavity, then its image is focused on a Charge Coupled Device (CCD) of a camera. By individually detecting the polarization of the separate parts of the photoluminescence (imaged light) on CCD, one may deduce the presence of entanglement between different spatial regions of the BEC as imaged on the CCD. (b) The enlarged image resolved from CCD. The middle dashed line shows the regions defining the two spin components used to detect entanglement.
			\label{experimentsetup} }
	\end{figure}

	In this paper, we propose a method of generating entanglement in a split polariton BEC and give an experimental scheme of detecting entanglement (see Fig. \ref{experimentsetup}). A single spinor polariton BEC is initially excited in the quantum wells (QWs) by optically pumping. Due to the natural self-interactions between the polaritons, this produces a one-axis twisting effect, producing multi-particle entanglement which involves all polaritons in the BEC. The single BEC is then spatially split into two ensembles which produces two separate spins. We note that this splitting procedure can be either a physical split or a virtual splitting procedure, where the image of the polaritons is partitioned into two (Fig. \ref{experimentsetup} (b)) \cite{yumang19}. After the splitting procedure, the sub-systems are still entangled due to the one-axis twisting producing multiparticle entanglement (see Fig. \ref{becsplit}) \cite{Tichy12,Bouvrie16,Bouvrie19}. We use a spin mapping to map our system with particle number fluctuations onto a fixed particle number space in order to use well-established spin correlators developed to detect entanglement. We calculate logarithmic negativity and correlation-based criteria to demonstrate that multi-particle entanglement exists not only in each BEC, but in a spatially separated configuration between two BECs. We show that our system exhibits stronger entanglement for larger particle number sectors in various regimes. By adjusting realistic system parameters one can improve the entanglement level.

	This paper is organized as follows. In Sec. \ref{ii} we introduce the theoretical model for a single spinor exciton-polariton condensate, and introduce the splitting operation, which produces two spatially separate BECs. In Sec. \ref{iii}, we numerically simulate our method and analyze our simulation results. In Sec. \ref{iv}, we show the main results of entanglement generation and detection by using different entanglement criteria. Finally, in Sec. \ref{v} we summarize and discuss our results.

	\section{Spin squeezed polariton condensates}\label{ii}
	
	\subsection{Theoretical model}
	
	We now describe the theoretical model used to simulate our interacting spinor polariton condensate. For further details we refer the reader to Ref.\cite{jingyan21}, which analyzes a similar situation prior to splitting. The master equation for the spinor polariton BEC is 

	\begin{align}	
		\frac{d\rho}{dt}= - \frac{i}{\hbar}[H_{\text{system}},\rho]-\frac{\gamma}{2}{\cal L} [a,\rho]-\frac{\gamma}{2} {\cal L} [b,\rho],
		\label{masterequation}
	\end{align}
	where the Hamiltonians $H_{\text{system}}=H_{0}+H_{\text{pump}}+H_{\text{int}}$ is defined 
	\begin{align}	
		H_{0} &=\hbar \Delta(a^{\dagger}a+b^{\dagger}b),\nonumber\\
		H_{\text{pump}} &=\hbar A(a^{\dagger}e^{-i\theta_{a}}+ae^{i\theta_{a}}+b^{\dagger}e^{-i\theta_{b}}+be^{i\theta_{b}}),\nonumber\\
		H_{\text{int}} &=\frac{\hbar U}{2}(a^{\dagger}a(a^{\dagger}a-1))+\frac{\hbar U}{2}(b^{\dagger}b(b^{\dagger}b-1))\nonumber\\
		&+\hbar V a^{\dagger}a b^{\dagger} b.
	\end{align}
	Here, $a^{\dagger},b^{\dagger}$ and $a,b$ are the creation and annihilation operators for the two zero momentum polariton spin species $ s = \pm 1 $ respectively, which obey bosonic commutation relations 
	\begin{align}
		[a,a^\dagger ]  & = [b, b^\dagger ] = 1, \nonumber \\
		[a,b] & =0 .
	\end{align} 

	The contribution of higher momentum polariton modes are not considered in our proposal as they do not affect the spin squeezing entanglement, which is the focus of this study. The above Hamiltoinian models resonant excitation, where the polaritons are typically excited at zero in-plane momentum, such that the remaining momenta are relatively unpopulated.  One may also consider off-resonant excitation, where other momenta will also be present, but in such a scheme only the zero momentum polaritons should be examined, which could be achieved by filtering in momentum space. We note that resonant excitation techniques have been used in numerous experimental studies of polariton BECs, and is considered to be an equivalent way of obtaining a condensed polariton cloud, although it lacks the condensation step that characterizes the BEC phase transition \cite{Adiyatullin17,Takesue04,Boulier14}. The Hamiltonian $H_{0}$ defines the energy $\hbar \Delta $ of zero-momentum polaritons with respect to the pump laser. $H_{\text{pump}}$ is the Hamiltonian for the pump laser with amplitude $A$, and $\theta_{a}$,$\theta_{b}$ represent the pumping phases of modes $a$ and $b$, respectively. The Hamiltonian $H_{\text{int}}$ includes the non-linear interaction energy $ \hbar U $ between the same spins and $ \hbar V $ for different spins. 
	The superoperator
	\begin{align}	
		{\cal L}[a,\rho]& =a^{\dagger}a\rho+\rho a^{\dagger}a-2a\rho a^{\dagger}, \nonumber\\
		{\cal L}[b,\rho]& =b^{\dagger}b\rho+\rho b^{\dagger}b-2b\rho b^{\dagger},
	\end{align}
	is the Lindbladian loss for photons leaking through the cavity. According to the master equation (\ref{masterequation}), the polariton population decays with rate $\gamma$. 
	\begin{figure}
		\includegraphics[width=\columnwidth]{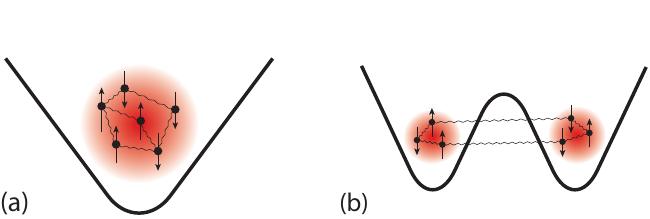}
		\caption{Entanglement in a split polariton condensate. (a) A single spinor polariton BEC first forms in the QWs, generating multi-particle entanglement at steady-state, represented by the wiggly lines. (b) The external potential trapping the condensate is modified such that it is spatially split into two BECs. The entanglement is transformed to a non-local form where it exists between the two split BECs.
			\label{becsplit} }
	\end{figure}

	To solve the master equation, we decompose the density matrix in the Fock basis, and numerically evolve the master equation. The density matrix can be written as
	\begin{align}	
		\rho=\sum_{klk'l'} \rho_{klk'l'}|k, l \rangle
		\langle k' ,l'|,
		\label{densmatexp}
	\end{align}
	where
	\begin{align}
		| k, l \rangle  =  \frac{(a^{\dagger}) ^{k} (b^{\dagger})^{l}}{\sqrt{k ! l !}} |0\rangle,
		\label{fockstatedef}
	\end{align}
	are the normalized Fock states that obey $\langle k,l|k',l'\rangle = \delta_{k k'}  \delta_{l l'}$.

	\subsection{Splitting the polariton condensate}
	
	Initially the spin modes $a$ and $b$ form a single BEC with all polaritons forming a multipartite entangled state due to the non-linear interaction as illustrated in Fig. \ref{becsplit} (a). In order to obtain the sub-modes $a_{1},a_{2},b_{1},b_{2} $  of the two spins $a$ and $b$, we apply the transformation 
	\begin{align}
		a \rightarrow \frac{1}{\sqrt{2}}(a_{1}+a_{2}),\nonumber\\
		b \rightarrow \frac{1}{\sqrt{2}}(b_{1}+b_{2}).\nonumber\\
		\label{split1}
	\end{align}
	The above splitting implies that there exists unoccupied modes undergoing the transformation
	\begin{align}
		\widetilde{a} \rightarrow \frac{1}{\sqrt{2}}(a_{1}-a_{2}),\nonumber\\
		\widetilde{b} \rightarrow \frac{1}{\sqrt{2}}(b_{1}-b_{2}).\nonumber\\
		\label{split2}
	\end{align}
	This transformation corresponds to a coherent splitting process similar to that shown in Fig. \ref{becsplit} (b). Alternatively, it could correspond to the virtual splitting as that shown in Fig. \ref{experimentsetup}, where the polariton condensate is split into two parts according to two spatial regions. These spatial regions have a one-to-one relation to the optical modes that emerge from the microcavity and hence may be spatially imaged according to the scheme shown in Fig. \ref{experimentsetup} (b). The above splitting operation forms either two physically separate BECs or two distinct halves of a BEC, and changes the entanglement structure, which we show in Fig. \ref{becsplit} (b). After the split, the Fock states transform as

	\begin{align}
		|k, l \rangle \rightarrow & \frac{1}{\sqrt{k ! l !}}
		\left(\frac{a_{1}^{\dagger}+a_{2}^{\dagger}}{\sqrt{2}}\right)^{k}
		\left(\frac{b_{1}^{\dagger}+b_{2}^{\dagger}}{\sqrt{2}}\right)^{l}  |0\rangle \nonumber\\
		=&\frac{1}{\sqrt{2^{k+l}} \sqrt{k ! l !}} 
		\sum_{nm} {k \choose n} {l \choose m} \nonumber\\
		&\times (a_{1}^{\dagger})^{n} (a_{2}^{\dagger})^{k-n} (b_{1}^{\dagger})^{m} (b_{2}^{\dagger})^{l-m} |0\rangle  \nonumber\\
		=&\frac{1}{\sqrt{2^{k+l}}}  \sum_{nm} \sqrt{ {k \choose n} {l \choose m} } |n,m,k-n,l-m \rangle,
		\label{11111}
	\end{align}
	where the normalized Fock state with four modes can be written as 
	\begin{align}
		|k_1,l_1,k_2,l_2\rangle = \frac{(a_{1}^{\dagger})^{k_1} 
			(b_{1}^{\dagger})^{l_1}(a_{2}^{\dagger})^{k_2} (b_{1}^{\dagger})^{l_2}}{\sqrt{k_1 ! l_1 ! k_2 ! l_2 ! }}|0\rangle.
	\end{align}
	Substituting the above into (\ref{densmatexp}), the density matrix of the split condensate is written in general as
	\begin{align}	
		\rho^{\text{sp}} =& \sum_{\substack{kl\\k'l'}}  \sum_{\substack{nm\\n'm'}} \frac{\rho_{klk'l'}}      {\sqrt{2^{k+l+k'+l'}}}   \sqrt{{k \choose n}     {l \choose m} {k' \choose n'} {l' \choose       m'}}     \nonumber\\     
		&\times |n,m,k-n,l-m  \rangle
		\langle  n',m',k'-n',l'-m'|.
		\label{densmatexpnew}
	\end{align}

	The spin operators on the split BEC are defined as

	\begin{align}
		S_{j}^{x} &=a^{\dagger}_{j} b_{j}+b_{j}^{\dagger}a_{j}, \nonumber\\
		S_{j}^{y} &=i(b^{\dagger}_{j} a_{j}-a^{\dagger}_{j} b_{j}), \nonumber\\
		S_{j}^{z} &=a^{\dagger}_{j} a_{j}-b^{\dagger}_{j} b_{j},
		\label{schwingerspin}
	\end{align}
	where $j \in \{1,2 \}$ labels the two BECs (either physical or virtual). These spin operators obey bosonic commutation relations

	\begin{align}
		[S^{l}, S^{m}]=2i\epsilon_{lmn}S^{n},
	\end{align}
	where $\epsilon_{lmn}$ is the Levi-Civita symbol and $ l,m,n \in  \{x,y,z\} $. The number operators for the two parts can be written as

	\begin{align}
		{\cal N}_j = a_j^\dagger a_j + b^\dagger_j b_j,
		\label{numberoperator}
	\end{align}
	where $j \in \{1,2 \}$.

	\subsection{Number fixing}

	The exciton-polariton BEC system is an open dissipative system and does not obey conservation of total polariton number. In context of atomic BECs, the total atom number $N$ is assumed to be fixed for a single run of the experiment. Any relation that is derived for fixed atom number (such as entanglement criteria) is not necessarily valid if the total particle number fluctuates.  In order to deal with this, we thus use a similar approach to Ref.\cite{jingyan21}  (Sec. \uppercase\expandafter{\romannumeral2}) to map $\rho^{\text{sp}}$ onto a fixed Hilbert space. Thus we define the density matrix in the $N$-sector as 

	\begin{align}	
		\rho^{\text{sp}}_N=\frac{\Pi_N \rho^{\text{sp}} \Pi_N}{p_{N}},
		\label{rhospfixed}
	\end{align}
	where 

	\begin{align}
		\Pi_N = &\sum_{N_{1}=0}^N \sum_{k_{1}=0}^{N_1} \sum_{k_{2}=0}^{N-N_1} |k_{1}, N_{1}-k_{1}, k_{2}, N-N_{1}-k_{2} \rangle \nonumber\\
		& \times \langle k_{1}, N_{1}-k_{1}, k_{2}, N-N_{1}-k_{2} | 
		\label{spinprojector}
	\end{align}
	is the projector on the $N-$particle subspace, and $N_{1}$ is the number of polaritons of the first BEC. The probability of the $N$-sector is defined as

	\begin{align}	
		p_{N}=\text{Tr}(\Pi_ N \rho^{\text{sp}} \Pi_N),
	\end{align}
	which satisfies the relation

	\begin{align}	
		\sum_{N} p_{N}=1.
	\end{align}
	Next we define the expectation values of quantum operator  $  {\cal O } $ in fixed $N$-sectors

	\begin{align}
		\langle {\cal O } \rangle_N \equiv \text{Tr} ( \rho_{N}  {\cal O } ) , 
		\label{expectation}
	\end{align}
	where $\rho_N$ is the projection of $\rho$ in a fixed $N$ space and the subscript $N$ refers to the fixed subspace. Therefore, the total polariton number would be
	\begin{align}	
		\langle {\cal N}_1 \rangle_N +\langle {\cal N}_2 \rangle_N  = N.
	\end{align}
	The variance of operator ${\cal O}$ for $N$-sector is defined as
	\begin{align}	
		\text{Var}_N ({\cal O})= \langle {\cal O}^2 \rangle_N - \langle {\cal O} \rangle_{N}^2.
	\end{align}

	The projector (\ref{spinprojector}) involves a fixed polariton number $N$. However, the total polariton number collapses to a fixed $N_1$ and $N_2$ after measurement. To define the projector on the fixed $N_1,N_2$ space $(N_2=N-N_1)$, we denote 

	\begin{align}
		\Pi_{N_1,N_2} =  &\sum_{k_{1}=0}^{N_1} \sum_{k_{2}=0}^{N_2} |k_{1}, N_{1}-k_{1}, k_{2}, N_2-k_{2} \rangle \nonumber\\
		& \times \langle k_{1}, N_{1}-k_{1}, k_{2}, N_2-k_{2} | ,
		\label{spinprojectorfixeN1}
	\end{align}
	which gives a fixed particle number on two halves. Thus the expectation values for the operator $\cal O$ in this space can be written as

	\begin{align}
		\langle {\cal O } \rangle_{N_1,N_2} \equiv \text{Tr} ( \rho_{N_1,N_2}  {\cal O } ) , 
		\label{expectationfixed}
	\end{align}
	we then obtain the relation of the expectation value of $\cal O$ for the two types of number fixing:

	\begin{align}
		\langle {\cal O } \rangle_N
		&= \text{Tr}(\rho_N {\cal O}) \nonumber\\
		&= \sum_{N_1=0}^N \sum_{N'_1=0}^N \text{Tr}(\Pi_{N_1,N_2} \rho_N \Pi_{N'_1,N'_2} \cal O) \nonumber\\
		&= \sum_{N_1=0}^N p_{N_1,N_2|N}\text{Tr} ( \rho_{N_1,N_2}  \cal O)\nonumber\\
		&= \sum_{N_1=0}^N p_{N_1,N_2|N} \langle {\cal O} \rangle_{N_1,N_2},
		\label{representation}
	\end{align}
	where $p_{N_1,N_2|N}$ is the conditional probability satisfies $\sum_{N_1}^N p_{N_1,N_2|N} = 1$, we assume $\cal O$ is a locally particle number conserving operator, and we use the fact that $\Pi_{N}^2=\Pi_N, \Pi_{N_1,N_2}^2=\Pi_{N_1,N_2}$. The above relations will be useful when it comes to examining correlation-based entanglement detection criteria, since these are often derived in the context of fixed $N_1, N_2 $ and we wish to relate these to number fluctuating averages.

	\section{Numerical simulation}\label{iii}
	
	\subsection{Evaluation of expectation values}
	
	In simulating the master equation (\ref{masterequation}), a truncation is necessary, since the full Hilbert space is unbounded. Therefore we impose a cutoff $N_{\text{max}}$, which means that the number of bosons that occupy each mode is restricted to $ k,l \in [0,N_\text{max} ] $. Any states with $ k,l>N_\text{max}$ are set to have zero amplitude. We note that the calculation of the effective spin still involves consideration of the truncation space within its context \cite{jingyan21}. 
	
	We then use (\ref{spinprojector}) to project the states on fixed total number $N$. We note that physically such a projection is automatically done when any measurement is performed. In any entanglement detection procedure, one requires detection of correlation between the two halves of the condensate. This involves detecting polaritons on the two sides of the condensate, and implicitly this involves a number fixing procedure. The density matrix (\ref{densmatexpnew}) is defined in a large Hilbert space with four spin modes $a_1,b_1,a_2,b_2$. Due to the numerical overhead with directly calculating the split four mode case, we calculate the expectation value of spin quantities $\cal O$ based on the original space before the splitting transformation, which contains only two modes. For example, the spin operators under this transformation will be written as

	\begin{align}	
		S^x_j = a^{\dagger}_j b_j + b^{\dagger}_j a_j 
		&\rightarrow \frac{1}{2} ( a^{\dagger} b + a^{\dagger} \widetilde b + {\widetilde a}^{\dagger} b + {\widetilde a}^{\dagger} \widetilde b) \nonumber \\
		&+  \frac{1}{2} (b^{\dagger} a + b^{\dagger} \widetilde a + {\widetilde b}^{\dagger} a+ {\widetilde b}^{\dagger} \widetilde a ), \nonumber \\     
		S^y_j = i(b^{\dagger}_j a_j - a^{\dagger}_j b_j) 
		&\rightarrow \frac{i}{2} ( b^{\dagger} a + b^{\dagger} \widetilde a + {\widetilde b}^{\dagger} a + {\widetilde b}^{\dagger} \widetilde a) \nonumber \\
		&-  \frac{i}{2} (a^{\dagger} b + a^{\dagger} \widetilde b + {\widetilde a}^{\dagger} b+ {\widetilde a}^{\dagger} \widetilde b ), \nonumber \\   
		S^z_j = a^{\dagger}_j a_j - b^{\dagger}_j b_j 
		&\rightarrow \frac{1}{2} ( a^{\dagger} a + a^{\dagger} \widetilde a + {\widetilde a}^{\dagger} a + {\widetilde a}^{\dagger} \widetilde a) \nonumber \\
		&-  \frac{1}{2} (b^{\dagger} b + b^{\dagger} \widetilde b + {\widetilde b}^{\dagger} b+ {\widetilde b}^{\dagger} \widetilde b ), \nonumber \\
	\end{align}
	where $j \in \{1,2 \}$, and we applied the inverse unitary transformation of the splitting procedure 

	\begin{align}
		a_1 \rightarrow \frac{1}{\sqrt{2}}(a+\widetilde a), \nonumber\\
		b_1 \rightarrow \frac{1}{\sqrt{2}}(b+\widetilde b), \nonumber\\ a_2 \rightarrow \frac{1}{\sqrt{2}}(a-\widetilde a), \nonumber\\b_2 \rightarrow \frac{1}{\sqrt{2}}(b-\widetilde b). 
		\label{inversetran}
	\end{align}
	The transformed spin operators involve both the original modes $ a,b $ as well as the unoccupied modes $ \widetilde{a}, \widetilde{b} $. Since we know that prior to the splitting operations $\widetilde a,\widetilde b$ annihilation operators are unoccupied, expectation values involving the operators $ \widetilde{a}, \widetilde{b} $ will give zero. For example, expectation values of the local modes give
	\begin{align}
		\langle S^x_j \rangle_N = \frac{1}{2} \langle a^{\dagger} b + b^{\dagger} a \rangle_N, \nonumber\\
		\langle S^y_j \rangle_N = \frac{i}{2} \langle b^{\dagger} a - a^{\dagger} b \rangle_N, \nonumber\\
		\langle S^z_j \rangle_N = \frac{1}{2} \langle a^{\dagger} a - b^{\dagger} b \rangle_N. \nonumber\\
	\end{align}
	where $j \in \{1,2 \}$. For second order spin correlations, we have
	\begin{align}
		&\langle S_{1}^{x}S_{2}^{x} \rangle_N \nonumber\\
		&= \frac{1}{4} \langle a^{\dagger} b a^{\dagger} b + a^{\dagger} b b^{\dagger} a - a^{\dagger} a +b^{\dagger} a  a^{\dagger} b -b^{\dagger} b + b^{\dagger} a  b^{\dagger} a \rangle_N, \nonumber\\
		&\langle S_{1}^{y}S_{2}^{y} \rangle_N \nonumber\\
		&= \frac{1}{4} \langle a^{\dagger} b b^{\dagger} a - a^{\dagger} a - a^{\dagger} b a^{\dagger} b - b^{\dagger} a  b^{\dagger} a + b^{\dagger} a  a^{\dagger} b - b^{\dagger} b \rangle_N, \nonumber\\
		&\langle S_{1}^{z}S_{2}^{z} \rangle_N \nonumber\\
		&= \frac{1}{4} \langle a^{\dagger} a a^{\dagger} a - a^{\dagger} a -a^{\dagger} a  b^{\dagger} b -b^{\dagger} b  a^{\dagger} a +b^{\dagger} b b^{\dagger} b - b^{\dagger} b \rangle_N, \nonumber\\
	\end{align}
	where we used the commutation relation $[\widetilde a,\widetilde {a}^{\dagger}] = [\widetilde b,\widetilde {b}^{\dagger}]=1$.
	
	The elements of density matrix (\ref{densmatexpnew}) can also be obtained from the original space, which can be calculated by
	\begin{align}
		& \langle k_1,l_1,k_2,l_2| \rho^{\text{sp}} |k'_1,l'_1,k'_2,l'_2\rangle \nonumber\\
		=& \frac{1}{\sqrt{k_1 ! l_1 ! k_2 ! l_2 ! k'_1 ! l'_1 ! k'_2 ! l'_2 ! }} \nonumber\\
		&\times \langle 0|a_{1}^{k_1} 
		b_{1}^{l_1} a_{2}^{k_2} b_{1}^{l_2} \rho^{\text{sp}} (a_{1}^{\dagger})^{k'_1} 
		(b_{1}^{\dagger})^{l'_1}(a_{2}^{\dagger})^{k'_2} (b_{1}^{\dagger})^{l'_2} |0 \rangle \nonumber\\
		\rightarrow& \frac{1}{\sqrt{k_1 ! l_1 ! k_2 ! l_2 ! k'_1 ! l'_1 ! k'_2 ! l'_2 ! }}  \times \frac{1}{\sqrt{2}^{k_1+l_1+k_2+l_2+k'_1+l'_1+k'_2+l'_2}}\nonumber\\
		&\times \langle 0| a^{k_1+k_2} 
		b^{l_1+l_2} \rho (a^{\dagger})^{k'_1+k'_2} 
		(b^{\dagger})^{l'_1+l'_2} |0 \rangle, \nonumber\\
	\end{align}
	where again we used the inverse unitary transformation (\ref{inversetran}).
	
	In the non-interacting limit $(U/\gamma=V/\gamma=0)$, each $N$-sector corresponds to a spin coherent state $|1/\sqrt{2},1/\sqrt{2}\rangle \rangle_1 \otimes |1/\sqrt{2},1/\sqrt{2}\rangle \rangle_2$ in the pump regime where $\theta_a = \theta_b = 0$ and $\Delta = 0$ after spin mapping \cite{jingyan21}, where we define the spin coherent state

	\begin{align}
		| \alpha, \beta \rangle \rangle & = \frac{1}{\sqrt{N!}} ( \alpha a^\dagger + \beta b^\dagger)^N | 0 \rangle \nonumber \\
		& =    	\sum_k \sqrt{N \choose k} \alpha^k \beta^{N-k} | k, N-k \rangle.
		\label{spincohdef}
	\end{align}
	For the initial state that is polarized in the $S^{x}$ direction, we evaluate that 

	\begin{align}	
		\langle S_{1}^{x} \rangle_N =\frac{1}{2^N} \sum_{N_1} {N \choose N_{1}} N_1 = \frac{N}{2},
	\end{align}
	\begin{align}	
		\langle S_{2}^{x} \rangle_N =\frac{1}{2^N} \sum_{N_2} {N \choose N_{2}} N_2 = \frac{N}{2}.
	\end{align}
	Hence, for each chosen $N$-sector, we have the average number of polaritons for the two parts are $N/2$, which indicates the equivalence of the calculations in the large space and in the split procedure.
	
	\subsection{Effective entangling Hamiltonian}
	
	To show how the effect of one-axis twisting in this split procedure, we project the total spin operator $S^z$ on to the fixed $N_1,N_2$ space by using (\ref{spinprojector})

	\begin{align}	
		&\Pi_{N_1, N_2} S^z \Pi_{N_1, N_2} = \Pi_{N_1, N_2} a^{\dagger}a \Pi_{N_1, N_2}- \Pi_{N_1, N_2} b^{\dagger}b \Pi_{N_1, N_2} \nonumber\\
		&=\frac{1}{2}\Pi_{N_1, N_2}(a_{1}^{\dagger}a_{1}-b_{1}^{\dagger}b_{1} + a_{2}^{\dagger}a_{2}-b_{2}^{\dagger}b_{2})\Pi_{N_1, N_2} \nonumber\\ &+  \frac{1}{2}\Pi_{N_1, N_2}(a_{1}^{\dagger}a_{2}+ a_{2}^{\dagger}a_{1} - b_{1}^{\dagger}b_{2} -b_{2}^{\dagger}b_{1})\Pi_{N_1, N_2} \nonumber\\
		&\rightarrow \Pi_{N_1, N_2}(a_{1}^{\dagger}a_{1}-b_{1}^{\dagger}b_{1} + a_{2}^{\dagger}a_{2}-b_{2}^{\dagger}b_{2})\Pi_{N_1, N_2}\nonumber\\
		&=\Pi_{N_1, N_2}(S_1^{z}+S_2^{z})\Pi_{N_1, N_2},
	\end{align}
	where the cross terms return zero for a fixed number $N$. Using the above result we then obtain
	\begin{align}	
		&\Pi_{N_1, N_2} (S^z)^2 \Pi_{N_1, N_2} = (\Pi_{N_1, N_2} S^z \Pi_{N_1, N_2})^2 \nonumber\\
		&\rightarrow \Pi_{N_1, N_2}(S_1^{z}+S_2^{z})^2\Pi_{N_1, N_2} \nonumber\\ 
		&= \Pi_{N_1, N_2}((S_{1}^{z})^2+2S_{1}^{z}S_{2}^{z}+(S_{2}^{z})^2)\Pi_{N_1, N_2},
		\label{effHam}
	\end{align}
	where we applied the relations $\Pi_{N_1,N_2}^{2}=\Pi_{N_1,N_2}$ and $[\Pi_{N_1,N_2},S^z]=0$. Thus the effective spin of the squeezing generation operator $S_{z}^2$ on one single polariton BEC in a fixed $N_1,N_2$ subspace corresponds to the two spatial BECs case as $(S^{z})^2 \rightarrow  (S_{1}^{z})^2+2S_{1}^{z}S_{2}^{z}+(S_{2}^{z})^2$. This shows that we expect squeezing on each BEC individually due to the terms $ (S^z_1)^2 $ and $ (S^z_2)^2 $ and the term $2S_1^z S_2^z$ generates entanglement between two BECs. This is similar to the one-axis two spin (1A2S) squeezing Hamiltonian, which produces entanglement with a fractal time dependence  \cite{Byrnes13,Kurkjian13}.

	\begin{figure}
		\includegraphics[width=\columnwidth]{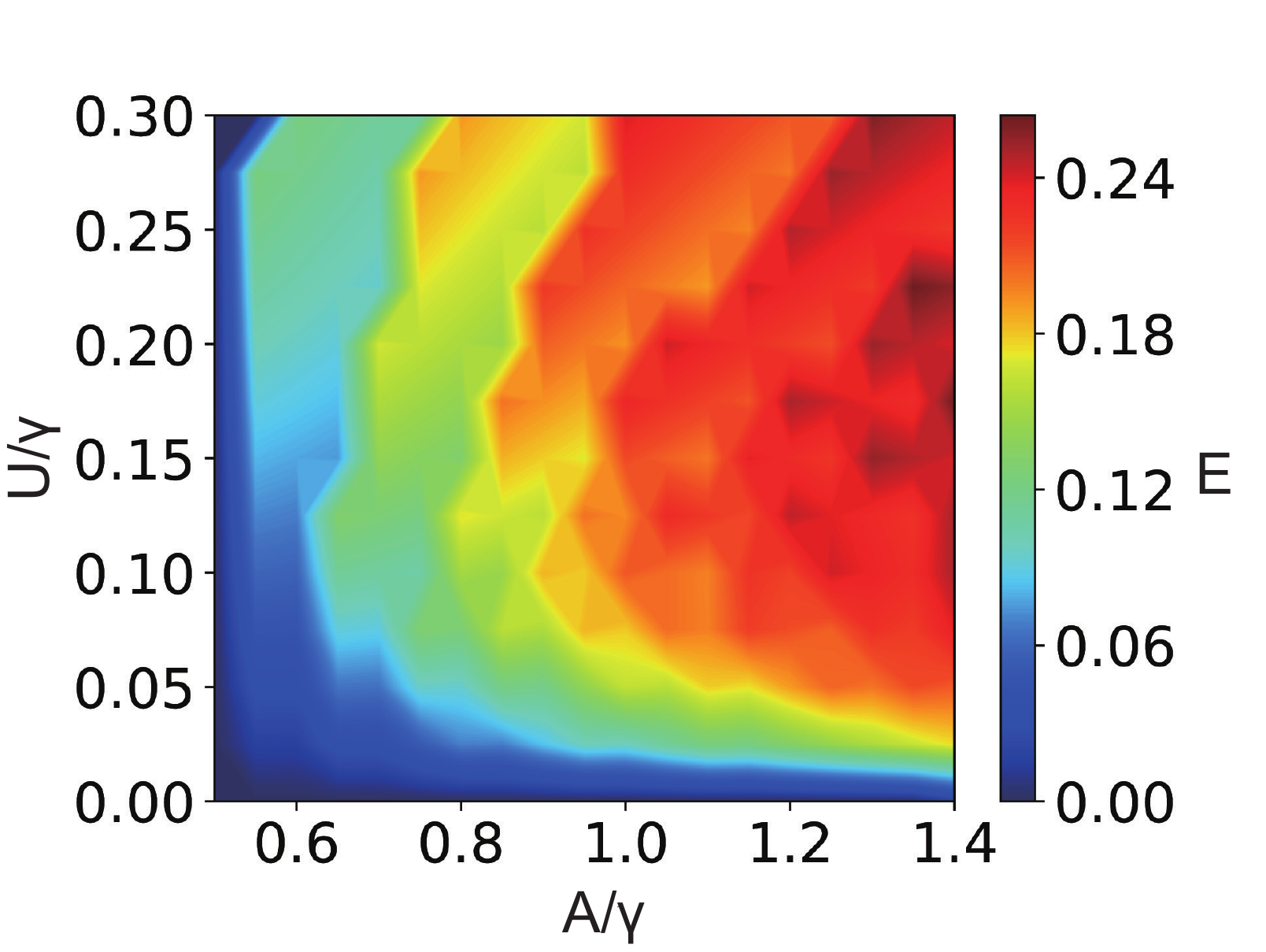}
		\caption{The logarithmic negativity (\ref{logcriteria}) as a function of pump rate $A/\gamma$ and the non-linear $S_{z}^2$ interaction parameter $U/\gamma$. Common parameters are $V/\gamma=0, \theta_a=\theta_b=0, N_\text{max}=10$.
			\label{log} }
	\end{figure}

	\section{Entanglement detection}\label{iv}
	
	\subsection{Logarithmic negativity}
	
	Logarithmic negativity is an entanglement monotone that is used to quantify the bipartite entanglement in mixed states \cite{Plenio05,Vidal02}. This is defined as

	\begin{align}
		E(\rho^{\text{sp}}_N)={\text{log}}_{2}||(\rho^{\text{sp}}_N)^{T_{2}}|| = {\text{log}}_{2}\sum_{i}|\lambda_{i}|,
		\label{logcriteria}
	\end{align}
	where $(\rho^{\text{sp}}_N)^{T_{2}}$ is the partial transpose of the second polariton BEC density matrix, $||{\cal X}||$ is the Schatten-1 norm of ${\cal X}$ and $|\lambda_{i}|$ is the absolute value of the eigenvalues of $(\rho^{\text{sp}}_N)^{T_{2}}$. The range of $E$ is from 0 to the maximum value $E_{\text{max}}={\text{log}}_2(N/2+1)$, where in the maximally entangled case $N_1=N_2=N/2$. We note that this result only involves the $N$-sector which has the maximum $p_{N}$ for the particular parameter set that we choose, i.e. the most likely measured $N$-sector. 
	
	In Fig.\ref{log}, we examine the logarithmic negativity, where we show (\ref{logcriteria}) as a function of the pumping rate $A/\gamma$ and the squeezing interaction parameter $U/\gamma$. We find that $E=0$ for a spin coherent state $(U/\gamma=0)$ and $E>0$ when both $U/\gamma, A/\gamma>0 $, as expected. The tendency of the growth of $E$ with $A/\gamma$ and $U/\gamma$ are clearly seen. We note that large $ N $ needs a smaller time to obtain the same squeezing level, as expected from the optimal squeezing time $ \propto 1/N^{2/3}$ for one-axis squeezing \cite{Kitagawa93,byrnes2020quantum}. Larger pumping and interaction corresponds to a higher level of squeezing, giving rise to more entanglement in our system. We thus expect that the entanglement should be present in the current polariton system at steady state for large pumping rate and high Q-cavity regime \cite{jingyan21}. Concretely, this would correspond to parameters corresponding to $1/\gamma>30 $ ps and $U/\gamma>0.3$. 

	\begin{figure}
		\includegraphics[width=\columnwidth]{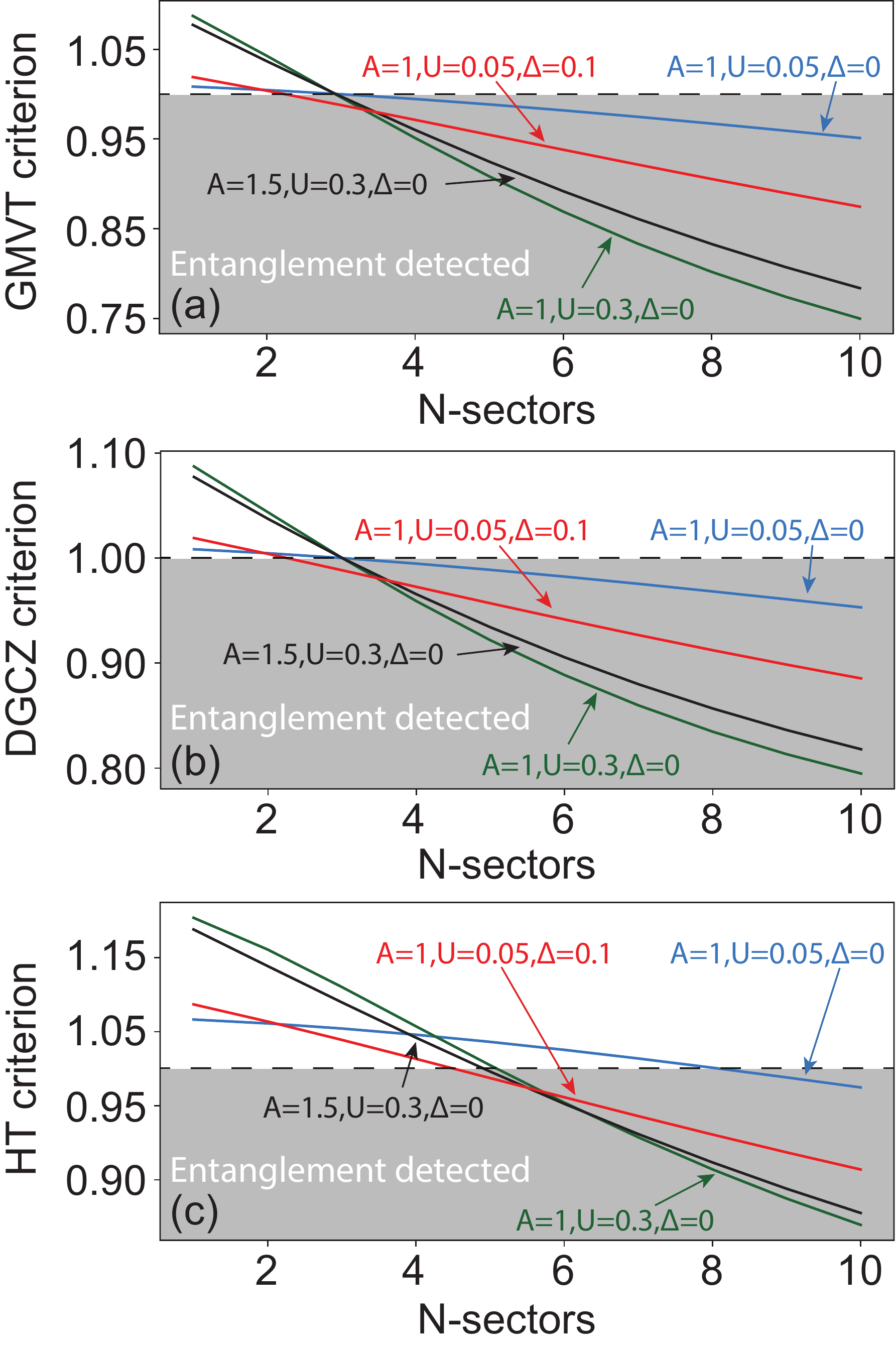}
		\caption{Entanglement criteria for the split polariton BECs system at steady-state. Criteria (\ref{GMVT}), (\ref{DGCZ}) and (\ref{Holfmann}) versus $N$-sectors are calculated in (a)-(c) respectively. The main experimental parameters $A/\gamma,U/\gamma,\Delta/\gamma$ are as marked. The shaded regions indicate the presence of entanglement. The indicated values are in units of $\gamma$. Common parameters are $V/\gamma=0, \theta_a=\theta_b=0, N_\text{max}=10$.
			\label{criteria} }
	\end{figure}

	\subsection{Correlation-based criteria}
	
	While a non-zero logarithmic negativity gives an unambiguous signal of entanglement, it may be difficult in practice to detect it in experiment due to the need for full density matrix tomography. Thus experimental limitations may require the use of alternative measures that are better suited to the available measurements. Several correlation-based entanglement detectors are available. The more sensitive detectors are the expectation values of total spin operators. The first one we consider is the Giovannetti-Mancini-Vitali-Tombesi (GMVT) criterion \cite{Giovannetti03}, which states that for any separable state

	\begin{align}
		\frac{\sqrt{\text{Var}_N (g_{y}S_{1}^y - S_{2}^y)\text{Var}_N (g_{z}S_{1}^z + S_{2}^z)}}{|g_{y}g_{z}|(|\langle S_{1}^x \rangle_N | +|\langle S_{2}^x \rangle_N |)} \ge 1,
		\label{GMVT}
	\end{align}
	where the $g_{y},g_{z}$ are free parameters to minimize the left hand side. The inequality (\ref{GMVT}) is true for all separable states. Hence a violation of the inequality indicates that the state must be entangled. In our case we choose $g_{y}=g_{z}=1$. The second criterion is the Duan-Giedke-Cirac-Zoller (DGCZ) criterion \cite{Duan00} valid for any separable state

	\begin{align}
		\frac{\text{Var}_N (S_{1}^y - S_{2}^y) + \text{Var}_N (S_{1}^z + S_{2}^z)}{2(|\langle S_{1}^x \rangle_N |+|\langle S_{2}^x \rangle_N |)} \ge 1.
		\label{DGCZ}
	\end{align}
	The third criterion is the Hofmann-Takeuchi (HT) criterion \cite{Hofmann03} valid for any separable state

	\begin{align}
		\frac{\text{Var}_N (S_{1}^x + S_{2}^x) + \text{Var}_N (S_{1}^y - S_{2}^y) + \text{Var}_N (S_{1}^z + S_{2}^z)}{2(\langle {{\cal N}_1} \rangle_N+ \langle {{\cal N}_2}\rangle_N)} \ge 1,
		\label{Holfmann}
	\end{align}
	where $N$ is the total polariton number of $N$-sector. The above inequalities have been converted from their fixed $N_1, N_2 $ relations to a fixed $ N $ through an averaging procedure. We note that the average variance of the operator ${\cal O}$ in the fixed $N_1, N_2$ space is either equal to or less than the variance defined using $N$-sectors, denoted by $\text{Var}_N ({\cal O})$ (see Appendix \ref{appendix}). Furthermore, the average expectation values are equal to $\langle {\cal O} \rangle_N$ (see Appendix \ref{appendixb}). Therefore, the violation of the inequalities (\ref{GMVT})-(\ref{Holfmann}) indicates the presence of entanglement within each $N$-sector.

	\begin{figure}
		\includegraphics[width=\columnwidth]{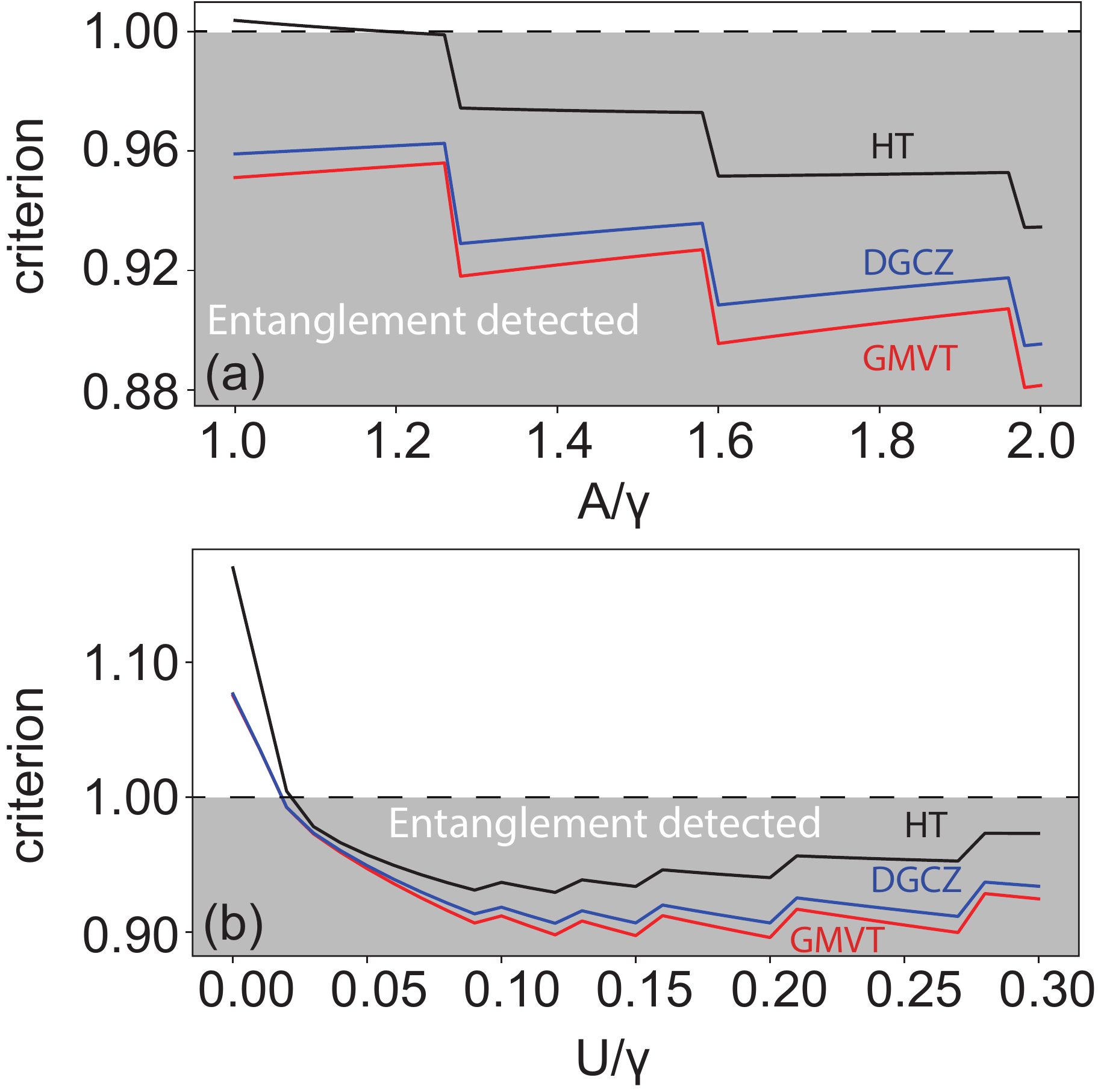}
		\caption{Entanglement criteria (\ref{GMVT}), (\ref{DGCZ}) and (\ref{Holfmann}) for the split polariton BECs system at steady-state as a function of (a) pump rate $A/\gamma$ and (b) interaction parameter $U/\gamma$. The three entanglement criteria are as marked. The $N$-sectors are chosen by the maximum $p_N$ for each parameter. The shaded regions indicate the presence of entanglement. Parameters are (a) $U/\gamma=0.3$, (b) $A/\gamma=1.5$. Common parameters are $\Delta/\gamma=0, V/\gamma=0, \theta_a=\theta_b=0, N_\text{max}=10$.
			\label{criteria2} }
	\end{figure}

	Fig. \ref{criteria}(a)-(c) shows the three criteria as a function of the $N$-sectors respectively. The first thing that we notice is the similar behavior of GMVT and DGCZ criteria.  The curves show that the former detects entanglement in a wider range than the latter. In Fig. \ref{criteria}(a), we see that for experimentally reasonable parameter choices, for larger $A/\gamma$ and small $\Delta/\gamma$, the entanglement criteria decrease monotonically with $N$. This corresponds to more squeezing for larger $N$, which was observed from the $Q$ functions and squeezing parameters in Ref. \cite{jingyan21}. Comparing the different parameters $U/\gamma$, we find that we obtain a higher entanglement level in a high Q-cavity, since the larger $U/\gamma$ represents more squeezing. Further, we show that the small detuning can enhance entanglement to a large extent. Depending upon the $ N $-sector examined, in some cases increasing the pump $ A/\gamma $ does not necessarily lead to an enhancement of the entanglement. We find a threshold in the $N$-sector, where below the threshold a larger pump rate $A/\gamma$ tends to increase entanglement, while above the threshold it decreases. For example, under HT criterion in Fig. \ref{criteria}(c), the threshold is $ N \approx 6 $.
	
	Fig. \ref{criteria2}(a) shows a ``staircase" dependence of GMVT, DGCZ and HT criteria. The staircase dependence is observed because we consider the most likely $ N $-sector to be measured, and with  increasing $ A/\gamma $ or $ U/\gamma $ this changes. For example, at $A/\gamma\sim1.25 $, the entanglement level suddenly decreases due to the change in the $N$-sector, then the entanglement level slightly reduces. What this shows is that while increasing $ A/\gamma $ can slightly degrade the entanglement within a fixed $ N $-sector, a larger pump can also change the most probable $ N $-sector, which can lead to an improvement of the entanglement. This is why in Fig. \ref{log} we generally observe an increase in entanglement with larger pumping. In experiments typically a larger pump rate is easily achieved, thus the most reachable regime is the higher $A/\gamma$. In Fig. \ref{criteria2}(b) we show these three criteria versus the interaction parameter $U/\gamma$. We see that the below the threshold $U/\gamma\sim0.09$, the entanglement level improves monotonically, but then saturates and again has a ``staircase" dependence due to the changes in $ N $-sector. Therefore, to obtain a higher entanglement level, a moderate interaction $ U /\gamma $ may be sufficient to obtain an optimized level of entanglement.

	\section{Conclusion}\label{v}
	
	In this work we theoretically proposed a method of generating spatially separated entanglement at steady-state in a spinor exciton-polariton BEC and gave two ways of realizing the experimental setup. In the first approach, the polaritons would be physically split in a coherent fashion, by raising an external potential, for example. The second approach involves virtually splitting the polariton condensate into two halves, by examining a spatially resolved near-field image of entanglement in the polariton BEC. In both of these approaches, equivalent results are obtained, in the ideal case. Technically, the virtual split is much easier to achieve than the physical split since such extra manipulations may involve additional sources of decoherence. However, the physical split is more in line with the notion of two separated entangled BECs, and is yet to be realized in any physical system. The initial formation of the condensate can be attributed to one-axis spin squeezing interaction between the polariton modes. This type of interaction leads to entanglement generation between all polaritons in the system. The formation of entanglement can be attributed to the cross term $2S_1^z S_2^z$ (\ref{effHam}). By examining and comparing the logarithmic negativity, GMVT, DGCZ, and HT criteria in various regimes, we show such entanglement can be detected between the two BECs. The entanglement can be improved with pump rate $A/\gamma$ increasing. We also find that a small detuning $\Delta/\gamma$ can enhance entanglement. Further, one may obtain an optimal entanglement level by adjusting the interaction parameter $U/\gamma$. To date, there has not been any report of entanglement generation within a polariton BEC. Several experiments in atomic BEC have been performed at the single BEC level to demonstrate entanglement, but two separate BECs have never been entangled. Due to the controllability of polariton condensate, there is an opportunity to experimentally realize some of these milestones in the near future.

	\begin{acknowledgments}
		This work is supported by the National Natural Science Foundation of China (62071301); NYU-ECNU Institute of Physics at NYU Shanghai; Shanghai Frontiers Science Center of Artificial Intelligence and Deep Learning; the Joint Physics Research Institute Challenge Grant; the Science and Technology Commission of Shanghai Municipality (19XD1423000,22ZR1444600); the NYU Shanghai Boost Fund; the China Foreign Experts Program (G2021013002L); the NYU Shanghai Major-Grants Seed Fund; Tamkeen under the NYU Abu Dhabi Research Institute grant CG008; and the SMEC Scientific Research Innovation Project (2023ZKZD55). 
		
	\end{acknowledgments}
	
	\appendix

	\section{Derivation of the variance average in the fixed $N_1,N_2$ space  }
	\label{appendix}
	
	The definition of the variance average of a quantum operator ${\cal O}$ is
	\begin{align}
		&\sum_{N_1=0}^N p_{N_1,N_2|N} \text{Var}_{N_1,N_2} ({\cal O})\nonumber\\ =& \sum_{N_1=0}^N p_{N_1,N_2|N} \langle {\cal O}^2 \rangle_{N_1,N_2} 
		- \sum_{N_1=0}^N p_{N_1,N_2|N} \langle {\cal O} \rangle_{N_1,N_2}^2.
	\end{align}
	Using Cauchy-Schwarz inequality \cite{Duan00}, we find
	\begin{align}
		&\sum_{N_1=0}^N p_{N_1,N_2|N} \text{Var}_{N_1,N_2} ({\cal O})\nonumber\\ \leq& \sum_{N_1=0}^N p_{N_1,N_2|N} \langle {\cal O}^2 \rangle_{N_1,N_2} 
		- \left(\sum_{N_1=0}^N p_{N_1,N_2|N} |\langle {\cal O} \rangle_{N_1,N_2}|\right)^2 \nonumber\\ \leq& \sum_{N_1=0}^N p_{N_1,N_2|N} \langle {\cal O}^2 \rangle_{N_1,N_2} 
		- \left(\sum_{N_1=0}^N p_{N_1,N_2|N} \langle {\cal O} \rangle_{N_1,N_2}\right)^2,       
	\end{align}
	by substituting $\langle {\cal O}^2 \rangle_{N_1,N_2}$,$\langle {\cal O} \rangle_{N_1,N_2}^2$ with $\langle {\cal O}^2 \rangle_N$,$\langle {\cal O} \rangle_N^2$ (\ref{representation}), we have

	\begin{align}
		\sum_{N_1=0}^N p_{N_1,N_2|N} \text{Var}_{N_1,N_2} ({\cal O})  \leq  \text{Var}_N ({\cal O}).
	\end{align}
	Replacing $\cal O$ with $S_1^x+S_2^x$,$S_1^y-S_2^y$,$S_1^z+S_2^z$ respectively, we obtain the relations in (\ref{GMVT})-(\ref{Holfmann}).

	\section{Derivation of the average of expectation values in the fixed $N_1,N_2$ space}
	\label{appendixb}
	
	The definition of the average of expectation values of a quantum operator ${\cal O}$ is

	\begin{align}
		\sum_{N_1=0}^N p_{N_1,N_2|N} \langle {\cal O} \rangle_{N_1,N_2},
	\end{align}
	by using the relations (\ref{representation}) we have

	\begin{align}
		\sum_{N_1=0}^N p_{N_1,N_2|N} \langle {\cal O} \rangle_{N_1,N_2} = \langle {\cal O} \rangle_N.
	\end{align}
	Replacing $\cal O$ with $S_1^x$,$S_2^x$,${{\cal N}_1}$,${{\cal N}_2}$ respectively, we obtain the relations in (\ref{GMVT})-(\ref{Holfmann}).

	% How to do the references:
	%% 1) First uncomment the below and compile
	\bibliographystyle{apsrev}
	\bibliography{ref}
	%% 2) Copy the .bbl file to below and comment out the above two lines.
	%\begin{thebibliography}{28}

\end{document}